\documentclass[final,1p,times]{elsarticle}

\usepackage{amsthm}
\usepackage{amsmath}
\usepackage{amssymb}
\usepackage{amsfonts}
\usepackage{graphicx}
\usepackage{color}
\usepackage{multirow}
\usepackage{pdflscape}

\usepackage{lipsum}
\makeatletter
\def\ps@pprintTitle{%
 \let\@oddhead\@empty
 \let\@evenhead\@empty
 \def\@oddfoot{}%
 \let\@evenfoot\@oddfoot}
\makeatother

\begin{document}

\begin{frontmatter}

\title{Systematic evaluation of a new combinatorial curvature for complex networks}

\author[label1]{R.P. Sreejith}
\author[label2,label3]{J\"urgen Jost}
\author[label2,label4]{Emil Saucan}
\author[label1]{Areejit Samal\corref{cor1}}
\address[label1]{The Institute of Mathematical Sciences, Homi Bhabha National Institute, Chennai, India}
\address[label2]{Max Planck Institute for Mathematics in the Sciences, Leipzig, Germany}
\address[label3]{The Santa Fe Institute, Santa Fe, New Mexico, USA}
\address[label4]{Department of Electrical Engineering, Technion, Israel Institute of Technology, Haifa, Israel}

\begin{abstract}
We have recently introduced Forman's discretization of Ricci curvature to the realm of complex networks. Forman curvature is an edge-based measure whose mathematical definition elegantly encapsulates the weights of nodes and edges in a complex network. In this contribution, we perform a comparative analysis of Forman curvature with other edge-based measures such as edge betweenness, embeddedness and dispersion in diverse model and real networks. We find that Forman curvature in comparison to embeddedness or dispersion is a better indicator of the importance of an edge for the large-scale connectivity of complex networks. Based on the definition of the Forman curvature of edges, there are two natural ways to define the Forman curvature of nodes in a network. In this contribution, we also examine these two possible definitions of Forman curvature of nodes in diverse model and real networks. Based on our empirical analysis, we find that in practice the unnormalized definition of the Forman curvature of nodes with the choice of combinatorial node weights is a better indicator of the importance of nodes in complex networks.
\end{abstract}

\end{frontmatter}

\section{Introduction}
\label{introduction}

Discrete mathematics, especially graph theory \cite{Harary1969,Bollobas1998}, has made vast contributions towards our understanding of the structure of complex networks \cite{Wasserman1994,Watts1998,Barabasi1999,Albert2002,Feng2007,Newman2010,Fortunato2010}. In network theory \cite{Wasserman1994,Watts1998,Barabasi1999,Albert2002,Feng2007,Newman2010,Fortunato2010}, a recent focus has been the development of measures inspired by geometry \cite{Eckmann2002,Lin2010,Lin2011,Narayan2011,Bauer2012,Jost2014,Ni2015,Sandhu2015a,Wu2015,Sreejith2016,Krioukov2016,Bianconi2016,Bianconi2017} for the characterization of complex networks. Curvature is a central concept in geometry, and in particular, Ricci curvature is a classical notion of Riemannian geometry that quantifies both volume growth of infinitesimal balls as well as dispersion rate of geodesics \cite{Jost2011}. Two discretizations of the classical Ricci curvature, Ollivier-Ricci curvature \cite{Ollivier2007,Ollivier2009,Ollivier2010,Ollivier2013} and Forman-Ricci curvature \cite{Forman2003}, have been introduced to the domain of complex networks \cite{Lin2010,Lin2011,Bauer2012,Jost2014,Ni2015,Sandhu2015a,Sreejith2016}. Firstly, Ollivier \cite{Ollivier2007,Ollivier2009,Ollivier2010,Ollivier2013} in 2007 proposed a discretization of the Ricci curvature. Subsequently, Ollivier-Ricci curvature was adapted to the setting of undirected graphs, and this concept has proven to be successful in the analysis of complex networks \cite{Lin2010,Lin2011,Bauer2012,Jost2014,Ni2015,Sandhu2015a}. Secondly, even before Ollivier, Forman \cite{Forman2003} had devised another discretization of the Ricci curvature. Recently, we \cite{Sreejith2016} have adapted the Forman-Ricci curvature to the domain of complex networks.

In the context of complex networks, the Forman curvature captures the second property of the Ricci curvature, namely, the dispersion rate of geodesics. This is achieved by adapting to the discrete setting of graphs, the classical Bochner-Weitzenb\"ock formula \cite{Jost2011}, which gives the connection between curvature and the Laplacian on a Riemannian manifold. The resulting definition of the Forman curvature of an edge \cite{Sreejith2016} is remarkably simple to compute in complex networks. Importantly, the definition of the Forman curvature \cite{Sreejith2016} captures on the one hand the combinatorial properties of the network and on the other hand naturally incorporates the weights of nodes and edges in the network. Thus, Forman curvature is suitable for the geometrical characterization of both weighted and unweighted networks. In our recent work \cite{Sreejith2016}, we have successfully shown that Forman curvature represents a natural, as well as computationally efficient tool for the analysis and classification of complex networks.

In contrast to node-based measures such as degree, clustering coefficient \cite{Holland1971,Watts1998} and betweenness centrality \cite{Freeman1977,Newman2010}, Forman curvature \cite{Sreejith2016} is an edge-based measure that quantifies the extent of spreading at the ends of edges in a network. The more is the spreading at the ends of edges, the more negative is the Forman curvature. We \cite{Sreejith2016} have previously shown that the Forman curvature is an indicator of the relative importance of edges in model and real networks. In contrast to several node-based measures, relatively few edge-based measures have been proposed to date for the analysis of complex networks. Besides Forman curvature \cite{Sreejith2016}, the available edge-based measures include edge betweenness \cite{Freeman1977,Brandes2001}, embeddedness \cite{Marsden1984} and dispersion \cite{Backstrom2014}. Edge betweenness \cite{Freeman1977,Brandes2001}, embeddedness \cite{Marsden1984} and dispersion \cite{Backstrom2014} have also been employed to quantify the importance of edges in complex networks, in particular, social networks. In this contribution, we therefore perform a comparative analysis of Forman curavture with edge betweenness, embeddedness and dispersion in model and real networks.

An attractive feature of the Forman curvature is its applicability to networks with any set of positive weights for nodes and edges \cite{Sreejith2016}. This feature renders the Forman curvature a powerful tool for analysis of weighted networks where the user can specify the best set of weights for nodes and edges. However, the available maps of real networks seldom specify the weights of nodes or edges in the network. While analyzing networks without prescribed weights, there are at least two possible choices for the node weights in the definition of the Forman curvature of an edge which are introduced in the next section. In this contribution, we also study in model and real networks the effect of the two possible choices of node weights in the definition of the Forman curvature of an edge.

Although Forman curvature is a measure associated to edges, it is also natural to desire an extension of the concept to nodes as most measures employed in network theory are associated to nodes in a network. However, intrinsically there isn't a unique manner to define the Forman curvature of a node based on the definition of the Forman curvature of an edge in a network. Instead, two natural ways present themselves to define the Forman curvature of a node based on the definition of the Forman curvature of an edge which are introduced in the next section. In this contribution, we also systematically examine in model and real networks the two definitions of the Forman curvature for nodes.

The remainder of this paper is organized as follows. In section \ref{theory}, we present a brief review of the Forman curvature, the definition of the Forman curvature of an edge, the two choices of node weights in the definition of the Forman curvature of an edge, and the two possible definitions of the Forman curvature of a node in a network. In section \ref{dataset}, we describe the model and real networks investigated here. In section \ref{results}, we present our results, and in section \ref{conclusion}, we conclude with a short summary and possible applications.


\section{Theory}
\label{theory}

\subsection{Forman curvature -- A brief overview}

Forman's \cite{Forman2003} discretization of the classical Ricci curvature is applicable to a large class of geometric objects, the so called (\textit{regular}) $CW$ \textit{complexes}. Forman's discretization of the Ricci curvature \cite{Forman2003} is derived on the basis of the so called Bochner-Weitzenb\"ock formula \cite{Jost2011} which in its classical form connects curvature and the Laplacian on a Riemannian manifold. However, while the well known version of the Bochner-Weitzenb\"ock formula relates the Laplacain operator of functions defined on a given manifold, a relatively less known though by no means any less important version of the Bochner-Weitzenb\"ock formula concerns {\it forms} rather than functions. It is this less well known variant of the Bochner-Weitzenb\"ock formula that enables adaptation to the large class of geometric objects, the so called $CW$ complexes, where cells play the role of the forms in the original, standard setting.

It is important to underline that the definition of the Forman curvature is valid for a wide spectrum of possible weights. In his original contribution, Forman \cite{Forman2003} had motivated the weights in his curvature definition from the point of view of simple geometrical quantities such as length, area and volume. However, the spectrum of possible sets of weights that are admissible in the definition of Forman curvature is far more extensive and general. An important if not the main motivation behind considering weighted manifolds (or, as in Forman's work, CW complexes), stems from the observation made by Cheeger, Gromov and others \cite{Gromov1999,Morgan2009} that to control collapse (degeneracy) of manifolds under curvature bounds (mainly, Ricci curvature bounds), one has to consider the volume and also more general measures. For other motivations, such as appertaining to minimal surfaces, we refer the reader to \cite{Morgan2005}.
While in several approaches, the considered measures typically satisfy certain smoothness properties, Forman's method allows for weights that do not satisfy any such properties. Indeed, Forman's approach can be viewed as an extreme discretization of the notion of metric measure space, where the underlying structure is \textit{well behaved} (a manifold), and the overlay measure also satisfies similar properties, by replacing the smooth manifolds with the more general $CW$ complexes along with not imposing any constraints on the attached discrete measure (i.e., the given weights). That being said, one of the remarkable characteristics of Forman's work is the fact that a geometric structure, so to say, can be recovered from any set of weights. More precisely, any given set of (positive) weights involved in the computation of the Ricci curvature, can be arbitrarily well approximated by a set of \textit{natural} or \textit{geometric} weights, i.e., weights that conform to dimensional scaling properties similar to those of the truly geometric measures such as length, area and volume.

\subsection{Forman curvature of an edge}

As classical Ricci curvature operates directionally along the vectors, the concept is naturally defined for edges in the network. Although for the general $n$-dimensional case, the Bochner-Weitzenb\"ock formula and the curvature term is given by a quite complicated formula, in the limiting 1-dimensional case of graphs or networks, the mathematical formula for the Forman curvature of an edge $e$ in the network, as given by \cite{Sreejith2016}, is quite simple:
\begin{equation}
\label{FormanRicciEdge}
\mathbf{F}(e) = w_e \left( \frac{w_{v_1}}{w_e} +  \frac{w_{v_2}}{w_e}  - \sum_{e_{v_1}\ \sim\ e,\ e_{v_2}\ \sim\ e} \left[\frac{w_{v_1}}{\sqrt{w_e w_{e_{v_1} }}} + \frac{w_{v_2}}{\sqrt{w_e w_{e_{v_2} }}} \right] \right)\,;
\end{equation}
where $e$ denotes the edge under consideration between two nodes $v_1$ and $v_2$, $w_e$ denotes the weight of the edge $e$ under consideration, $w_{v_1}$ and $w_{v_2}$ denote the weights associated with the nodes $v_1$ and $v_2$, respectively, $e_{v_1} \sim e$ and $e_{v_2} \sim e$ denote the set of edges incident on nodes $v_1$ and $v_2$, respectively, after excluding the edge $e$ under consideration which connects the two nodes $v_1$ and $v_2$. Note that the indices $e_{v_1} \sim e$ and $e_{v_2} \sim e$ below the summation sign on the right hand side of Eq. \ref{FormanRicciEdge} do not specify a double sum but rather specify a single sum, that is,
\begin{eqnarray*}
\sum_{e_{v_1}\ \sim\ e,\ e_{v_2}\ \sim\ e} \left[\frac{w_{v_1}}{\sqrt{w_e w_{e_{v_1} }}} + \frac{w_{v_2}}{\sqrt{w_e w_{e_{v_2} }}} \right]= \sum_{e_{v_1}\ \sim\ e} \frac{w_{v_1}}{\sqrt{w_e w_{e_{v_1}}}} + \sum_{e_{v_2}\ \sim\ e} \frac{w_{v_2}}{\sqrt{w_e w_{e_{v_2} }}}
\end{eqnarray*}

As we have have emphasized before, any collection of positive weights can be inputed in the formula for the Forman curvature of an edge, a proof of the flexibility and adaptiveness of Forman's curvature. Here, we would like to bring to the readers' attention the fact that, in practice, one might be confronted with two sets of weights that, in many cases, but by no means always, are identical. The first such set of weights are the ones that are to be inputed into the formula of Forman-Ricci curvature. These weights, however, do not necessarily coincide with the set of weights empirically determined for a specific real-world network. For instance, for certain problems, it is more natural not to consider the \textit{distance} between two nodes $v_1$ and $v_2$ in the network as the original weight $w_e$ associated to the edge $e$, but rather its inverse $\frac{1}{w_{e}}$ (see for example \cite{Saucan2005}). In fact, there exists a whole body of literature dedicated to the problem of determining \textit{the most expressive} metric corresponding to a certain type of network. As an example, we would like to refer the reader to one such classic work \cite{Rajaraman2002}. However, this represents more a problem of modeling a specific type of network or problem as exemplified by the two cases mentioned above. So, we concentrate below on the Forman curvature per se, and thus, also consider the weights as prescribed in the network.


\subsection{Two choices for node weights in the definition of Forman curvature}

The mathematical definition for the Forman curvature of an edge (Eq. \ref{FormanRicciEdge}) can give meaningful results for any set of positive node weights and positive edge weights. As noted before, Forman curvature \cite{Sreejith2016} is a simple yet powerful network measure that captures on one hand the topological structure of the network and on the other hand naturally encapsulates the prescribed weights of both nodes and edges in the network. In general, one expects weights to be associated with both nodes and edges in real networks, and this novel curvature measure has the innate ability to cope with such general weighing schemes. However, in practice, the currently available data for real networks rarely prescribe the node weights or edge weights in the network. Hence, while computing the Forman curvature of an edge using Eq. \ref{FormanRicciEdge} in networks without prescribed weights, there are two natural choices to assign node weights. The first choice is the combinatorial one where all nodes (as well as all edges) are assigned weight equal to 1. The second choice which might seem better in practice is to assign node weights equal to their degree.

The first combinatorial choice is the simplest one which has been widely used in the vast literature on topology and geometry of graphs, and because of that, in various applications in network theory. Moreover, and more pertinent to this study, this simple choice was also made by Forman and resides at the core of his first (unweighted) definition of Ricci curvature (See Definition 3 in \cite{Forman2003}). In addition, it is precisely this simple definition that allows Forman to obtain interesting and important results connecting the topology and geometry of a large class of spaces that includes triangular and tetrahedral meshes. Thus, we choose in our previous work \cite{Sreejith2016} node weight equal to 1 in the definition of the Forman curvature for the analysis of complex networks.

The second practical choice of using the node degree as node weight in the definition of the Forman curvature seems both an admissible and natural option. In fact, this choice also appears in the definition of the graph Laplacian \cite{Chung1997,Banerjee2009} which in turn suggests that this second option represents a strong counter-candidate for the combinatorial weight (i.e., node weight equal to 1). However, in stark contrast to the combinatorial weight, there are no significant theoretical results involving node degree as the node weight. Therefore, in this work, we perform a comparative analysis in model and real networks of the Forman curvature with the two choices of node weights in Eq. \ref{FormanRicciEdge}, i.e., node weight equal to 1 or node weight equal to degree.


\subsection{Normalized and Unnormalized Forman curvature of a node}

Based on the definition of the Forman curvature of edges (Eq. \ref{FormanRicciEdge}), there are two natural ways to define the Forman curvature of a node in a network. Firstly, the normalized Forman curvature of a node can be defined as the sum of the curvature of all edges incident on that node divided by the degree of that node, that is,
\begin{equation}
\label{NormalizedFormanNode}
\mathbf{F}(v) = \frac{1}{\text{deg}(v)}\sum_{e_v\ \sim\ v} \mathbf{F}(e_v) \,;
\end{equation}
where $\mathbf{F}(v)$ is the normalized Forman curvature of the node $v$, $\mathbf{F}(e_v)$ is the Forman curvature of edge $e_v$, $e_v \sim v$ denotes the set of edges incident on node $v$, and deg$(v)$ denotes the degree of node $v$. Secondly, the unnormalized Forman curvature of a node can be defined as the sum of the curvature of all edges incident on that node, that is,
\begin{equation}
\label{UnnormalizedFormanNode}
\mathbf{F}(v) = \sum_{e_v\ \sim\ v} \mathbf{F}(e_v) \,;
\end{equation}
where $\mathbf{F}(v)$ is the unnormalized Forman curvature of the node $v$, $\mathbf{F}(e_v)$ is the Forman curvature of edge $e_v$, and $e_v \sim v$ denotes the set of edges incident on node $v$.

In \cite{Sreejith2016}, we have analyzed diverse networks based on the normalized Forman curvature of a node (Eq. \ref{NormalizedFormanNode}). However, it is a priori unclear if normalized or unnormalized Forman curvature is better suited for the analysis of model and real networks. Our previous choice \cite{Sreejith2016} of the normalized Forman curvature was motivated by the properties of the normalized graph Laplacian \cite{Chung1997,Banerjee2009,Bauer2012,Mehatari2015} which is a standard tool for analysis of complex networks. While one can define an unnormalized graph Laplacian \cite{Chung1997}, the normalized graph Laplacian \cite{Chung1997,Banerjee2009,Bauer2012,Mehatari2015} was found to have better properties. Therefore, previous analytical work on graph Laplacian suggests that the normalized versions of network measures are preferable for the analysis of complex networks. In consequence, given the intrinsic connection between Laplacian and Forman curvature, we \cite{Sreejith2016} had decided to use the normalized Forman curvature of a node (Eq. \ref{NormalizedFormanNode}) to analyze complex networks. A similar normalized choice was also made by Ni \emph{et al} \cite{Ni2015} for the analysis of Ollivier-Ricci curvature in complex networks.

On the other hand, if one draws inspiration from concepts of Riemannian geometry, it is natural to choose the unnormalized Forman curvature of a node (Eq. \ref{UnnormalizedFormanNode}) to analyze complex networks. Indeed, this unnormalized choice, which is analogous to scalar curvature in Riemannian geometry \cite{Jost2011}, was also made by Sandhu \emph{et al} \cite{Sandhu2015a} for the analysis of Ollivier-Ricci curvature in complex networks. Thus, both normalized and unnormalized Forman curvature are legitimate choices for analysis of complex networks, well supported by theoretical justifications. Hence, in this work, we perform a comparative analysis of normalized and unnormalized Forman curvature in model and real networks.


\section{Model and real networks}
\label{dataset}

In this work, we have considered four benchmark models of complex networks: Erd\"{o}s-R\'{e}nyi (ER) \cite{Erdos1961}, Watts-Strogatz (WS) \cite{Watts1998}, Barab\'{a}si-Albert (BA) \cite{Barabasi1999} and Network geometry with flavor (NGF) \cite{Bianconi2016,Bianconi2017}. The ER model is the benchmark to generate random networks. The ER model gives an ensemble of random graphs $G(n,p)$ based on two parameters, the number of nodes $n$ and the probability $p$ that there exists an edge between any pair of nodes in the network. The WS model is widely used to generate small-world networks which have high clustering coefficient and small average path length. In the WS model, an initial network is generated with $n$ nodes on a regular ring lattice wherein each node is connected to its $k$ nearest neighbours. Thereafter, in the WS model, the endpoint of each edge in the regular ring lattice or initial network is rewired with probability $\beta$ to a new node wherein the new node is selected from all possible nodes in the network with uniform probability. The BA model is commonly used to generate scale-free networks which have power-law degree distribution. In the BA model, the initial network consists of $m_0$ nodes. Thereafter, in the BA model, the network grows in a step-wise manner through addition of one new node at a time. In the BA model, at each step a new node is added to the network with connections to $m$ $\le$ $m_0$ existing nodes such that the probability $p_i$ that the new node will have an edge to existing node $i$ is given by $$p_i = \frac{k_i}{\sum_j k_j},$$ where $k_i$ is the degree of existing node $i$ and the sum in the denominator is taken over the degree of all existing nodes $j$. Thus, the BA model implements a preferential attachment scheme wherein high-degree nodes are likely to acquire more edges over time compared to low-degree nodes. The NGF model can be used to generate networks with geometric structure and can display emergent hyperbolic geometry \cite{Bianconi2016,Bianconi2017}. NGF with $s=-1,0,1$ describe growing simplicial complexes defined in arbitrary dimension $d$. Generation of a NGF with dimension $d$ starts with a $d$-dimensional simplex. Note that a $d$-dimensional simplex is a fully connected graph of ($d+1$) nodes, and a $d$-dimensional simplex has $\delta$ faces which are $\delta$-dimensional simplices formed by a subset ($\delta$+1) of ($d+1$) nodes. Initially, an energy, drawn from a (random) distribution, is associated to each node in the NGF which remains constant throughout network evolution. Moreover, each $\delta$ face in the NGF is also associated with an energy equal to the sum of the energy of the nodes that constitute that face. Thus, an energy is associated to each simplex (i.e., nodes, links, triangular faces, etc.) in the NGF. Importantly, the NGF model of growing simplicial complexes evolves with either preferential attachment ($s=1$) or without preferential attachment ($s=-1,0$) depending on the value of the flavor $s$. For example, in the NGF with flavor $s=1$ and dimension $d=2$, a new node is added preferentially at each step of the network evolution such that the new node has connections to two adjacent nodes in the existing network leading to formation of a triangle or 2-simplex.

In addition to model networks, we have also investigated several real networks including Adjective-Noun adjacency \cite{Newman2006}, Email communication \cite{Guimera2003}, Euro road \cite{Subelj2011}, Facebook \cite{Mcauley2012}, Gnutella \cite{Leskovec2007}, Hamsterster friendship, Human protein interactions \cite{Rual2005}, Jazz musicians \cite{Gleiser2003}, PDZ domain interactions \cite{Beuming2005}, US Power Grid \cite{Leskovec2007} and Yeast protein interactions \cite{Jeong2001}. The adjective-noun adjacency network consists of 112 nodes and 425 edges wherein the nodes are nouns or adjectives in the novel David Copperfield and the edges represent their presence in adjacent positions in the novel. The Email communication network consists of 1133 nodes and 5451 edges wherein the nodes are users at the University Rovira i Virgili in Spain and the edges represent direct communication between users. The Euro road network consists of 1174 nodes and 1417 edges wherein the nodes are cities in Europe and the edges represent roads in the international E-road network linking them. The Facebook network consists of 2888 nodes and 2981 edges wherein the nodes are users of the online social network Facebook and the edges represent friendship between users. The Gnutella network consists of 6301 nodes and 20777 edges wherein the nodes are hosts in Gnutella peer-to-peer file sharing network from August 2002 and the edges represent connections between hosts. The Hamsterster friendship network consists of 2426 nodes and 16631 edges wherein the nodes are users of hamsterster.com and the edges represent friendship or family links between users. The human protein interactions network consists of 3133 nodes and 6726 edges wherein the nodes are human proteins and the edges represent interactions between them. The Jazz musicians network consists of 198 nodes and 2742 edges wherein the nodes are musicians and the edges represent collaboration between them. The PDZ domain interactions network consists of 212 nodes and 244 edges wherein the nodes are proteins and the edges represent PDZ-domain mediated interactions between them. The US Power Grid network consists of 4941 nodes and 6594 edges wherein the nodes are generators or transformers or substations in the western states of USA and the edges represent power supply lines. The yeast protein interactions network consists of 1870 nodes and 2277 edges wherein the nodes are yeast proteins and the edges represent interactions between them. Most of the investigated real networks were downloaded from the KONECT \cite{Kunegis2013} database.


\section{Results and Discussion}
\label{results}

\subsection{Distribution of Forman curvature of edges}

\begin{figure}
\includegraphics[width=.5\columnwidth]{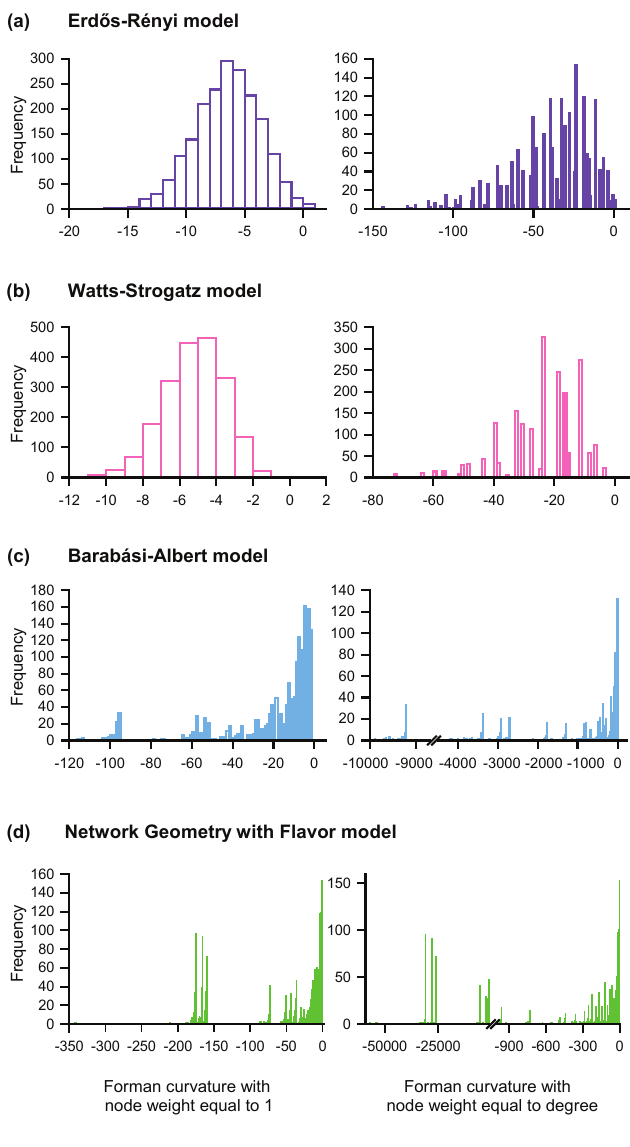}
\caption{Distribution in model networks of the Forman curvature of edges for the two choices of node weights. (a) Erd\"{o}s-R\`{e}nyi (ER) model. (b) Watts-Strogratz (WS) model. (c) Barab\`{a}si-Albert (BA) model. (d) Network geometry with flavor (NGF) model where flavor $s=1$ and dimension $d=2$.}
\label{dist_model_edge}
\end{figure}

\begin{figure}
\includegraphics[width=.5\columnwidth]{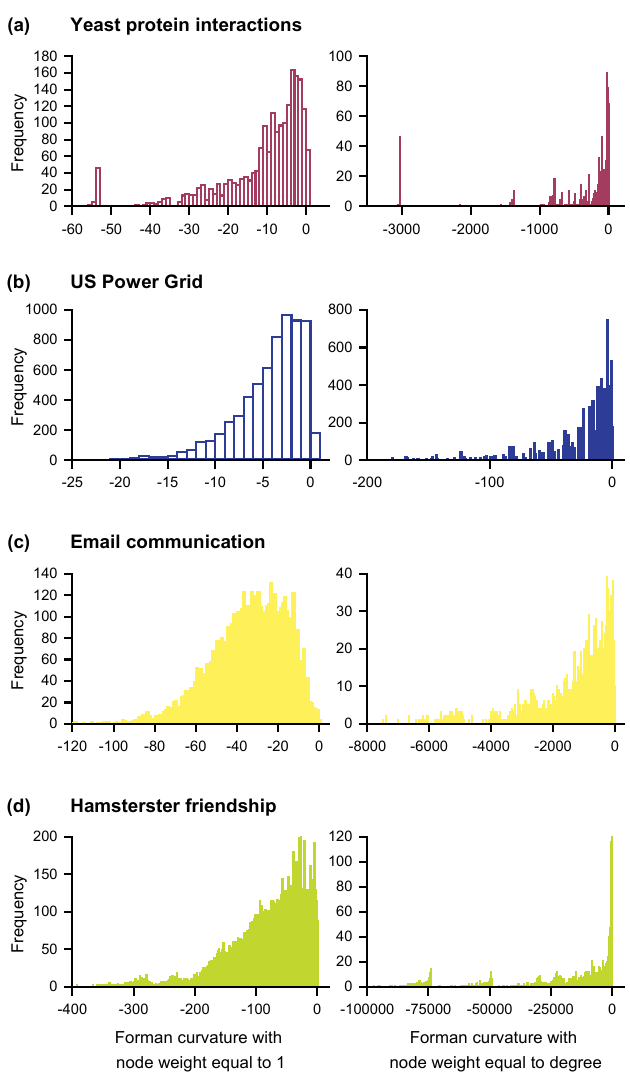}
\caption{Distribution in real networks of the Forman curvature of edges for the two choices of node weights. (a) Yeast protein interactions. (b) US Power Grid. (c) Email communication. (d) Hamsterster friendship.}
\label{dist_real_edge}
\end{figure}

\begin{figure}
\includegraphics[width=.7\columnwidth]{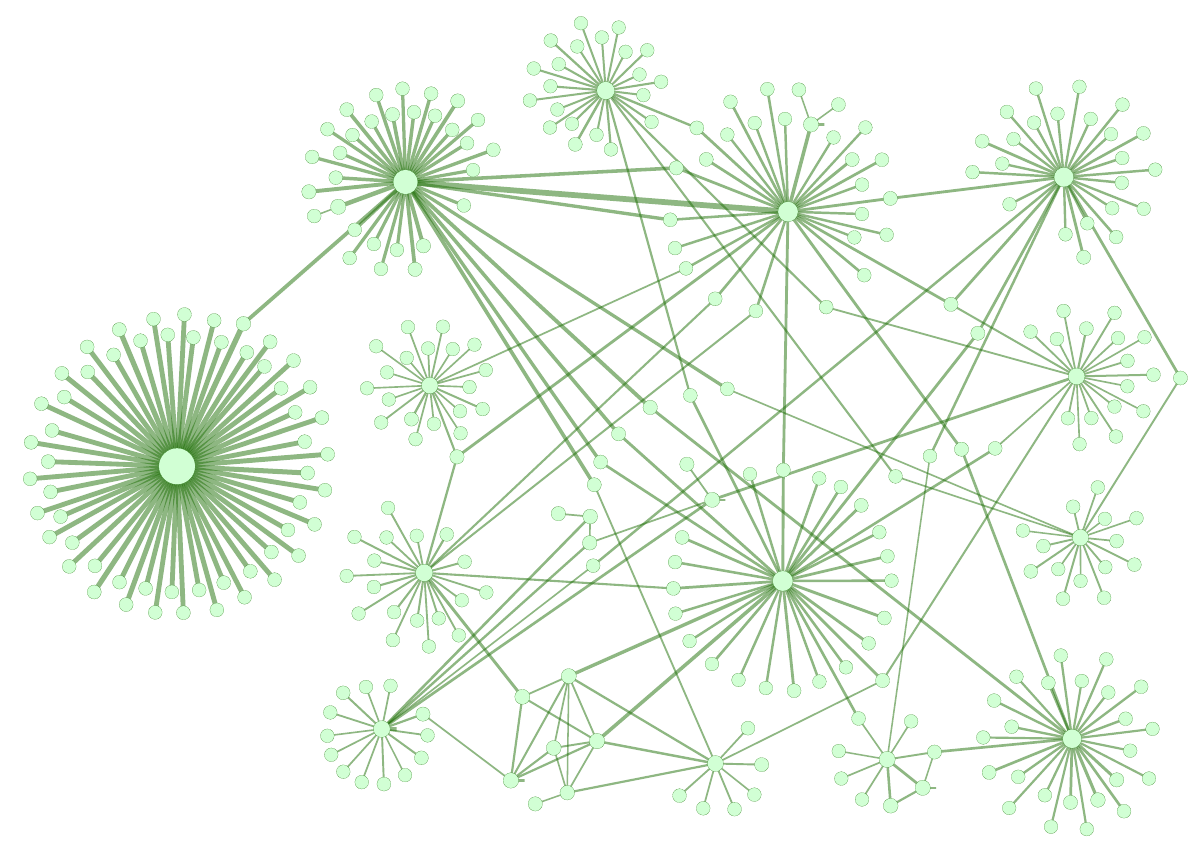}
\caption{Network visualization of the Yeast protein interactions. For this visualization, we sort edges in the ascending order of their Forman curvature with the choice of node weight equal to 1, and display only edges in the top 20\% of the sorted list. Width of edges is proportional to the absolute value of the Forman curvature with node weight equal to 1. Size of nodes is proportional to the unnormalized Forman curvature with node weight equal to 1.}
\label{yeastnet}
\end{figure}

We have investigated the distribution of the Forman curvature of edges for the two choices of node weights (see section \ref{theory}) in model and real networks. Obviously, the magnitude of the Forman curvature of edges with node weight equal to degree is higher than the Forman curvature of edges with node weight equal to 1 (Figures \ref{dist_model_edge} and \ref{dist_real_edge}). Among model networks, at a qualitative level, the nature of the distribution of the Forman curvature of edges for the two choices of node weights is the same for a given model network (Figure \ref{dist_model_edge}). Specifically, we find the distribution of the Forman curvature of edges to be narrow in random ER networks and small-world WS networks, while it is broad in scale-free BA networks and NGF network with $s=1$ and $d=2$ (Figure \ref{dist_model_edge}). Similar to model networks, at a qualitative level, the nature of the distribution of the Forman curvature of edges for the two choices of node weights is the same for a given real network (Figure \ref{dist_real_edge}). Figure \ref{yeastnet} visualizes a portion of the network of protein interactions in yeast where most edges have negative Forman curvature. This extends our previous results \cite{Sreejith2016} on the distribution of the Forman curvature of edges with the choice of node weight equal to 1 in model and real networks. Thus, the distribution of the Forman curavture of edges for the two choices of node weights can be used to classify different types of networks.


\subsection{Association of edge-based measures with Forman curvature of edges}

\begin{figure}
\includegraphics[width=.5\columnwidth]{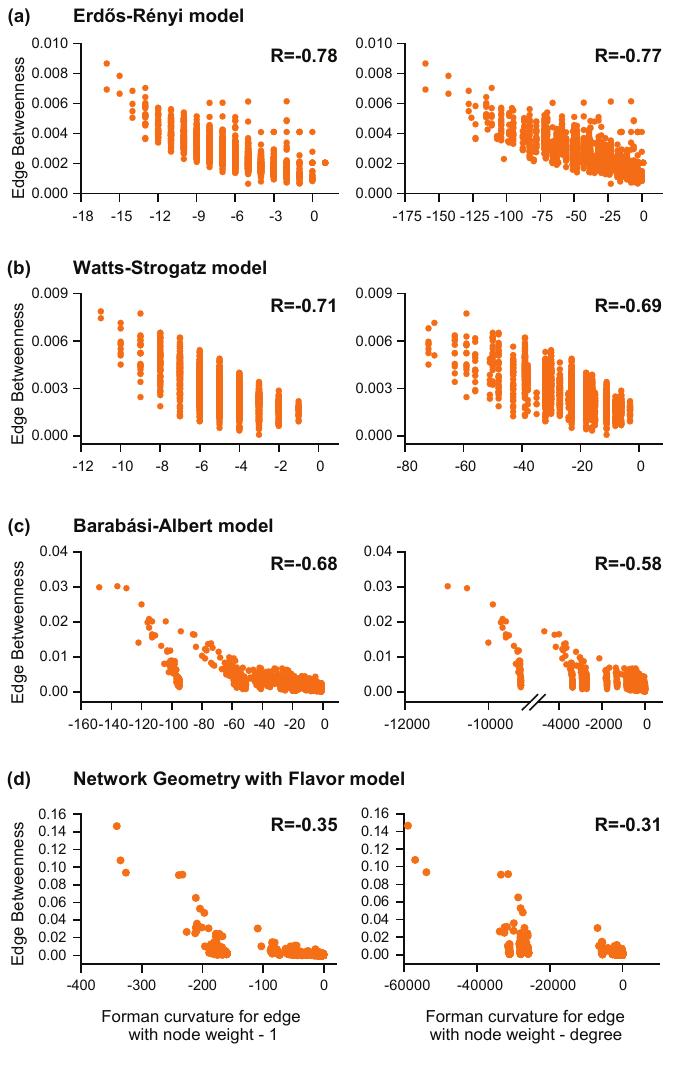}
\caption{Correlation between edge betweenness and the Forman curvature of edges for the two choices of node weights in model networks. (a) Erd\"{o}s-R\`{e}nyi (ER) model. (b) Watts-Strogratz (WS) model. (c) Barab\`{a}si-Albert (BA) model. (d) Network geometry with flavor (NGF) model where flavor $s=1$ and dimension $d=2$. We indicate the Pearson correlation coefficient $R$ in each plot.}
\label{corr_model_bet_edge}
\end{figure}

Edge betweennness centrality \cite{Freeman1977,Girvan2002,Newman2010} is a global measure which quantifies the number of shortest paths that pass through an edge in a network. An edge with high betweenness can be a bottleneck for flows in the network. We investigated the correlation between the two edge-based measures, edge betweenness and Forman curvature, in model and real networks (Figure \ref{corr_model_bet_edge};
Table \ref{table_edge_measures}). In model networks, we find that the Forman curvature of edges for the two choices of node weights has high negative correlation with edge betweenness in ER, WS and BA networks while a moderate negative correlation is observed in NGF networks (Figure \ref{corr_model_bet_edge}; Table \ref{table_edge_measures}). In real networks, we find that the Forman curvature of edges for the two choices of node weights has weak or no correlation with edge betweenness (Figure \ref{corr_model_bet_edge}; Table \ref{table_edge_measures}). Moreover, we find in both model and real networks that the correlation between the Forman curvature of edges with node weight equal to 1 and edge betweenness is similar to the correlation between the Forman curvature of edges with node weight equal to degree and edge betweenness (Figure \ref{corr_model_bet_edge}; Table \ref{table_edge_measures}). We should point out that the typical negative correlation noted above are a consequence of the choice of sign for Forman's curvature, a choice motivated by Riemannian Geometry (see also the discussion in the sequel). Thus, the correlations with the negative Forman curvature would be positive.

Embeddedness \cite{Marsden1984} is an edge-based measure that has been widely used to quantify the strength of ties in social networks. Embeddedness \cite{Marsden1984} of an edge in a network quantifies the number of neighbours that are shared by the two nodes anchoring the edge under consideration.  We also investigated the correlation between embeddedness and Forman curvature in model and real networks (Table \ref{table_edge_measures}). Among model networks, we find that embeddedness has no correlation with the Forman curvature of edges for the two choices of node weights in the ER and the WS models, while a significant negative correlation was obtained in the BA and NGF models (Table \ref{table_edge_measures}). Among real networks, we find that embeddedness has significant negative correlation with the Forman curvature of edges for the two choices of node weights in several networks, while no correlation was also obtained in multiple instances (Table \ref{table_edge_measures}). In fact, one may rather expect the edge-based measure, embeddedness, to be more related to the node-based measure, clustering coefficient, in complex networks.

Dispersion \cite{Backstrom2014} is an edge-based measure which was recently proposed to predict romantic relationships in social networks. In contrast to embeddedness, dispersion quantifies the extent to which the neighbours of the two nodes anchoring an edge are not themselves well connected \cite{Backstrom2014}. We also investigated the correlation between dispersion and Forman curvature in model and real networks (Table \ref{table_edge_measures}). Among model networks, we find that dispersion has no correlation with the Forman curvature of edges for the two choices of node weights in ER, WS and BA networks, while a significant negative correlation is seen in NGF networks (Table \ref{table_edge_measures}). Among real networks, we find that dispersion has no correlation with the Forman curvature of edges for the two choices of node weights in most of the considered networks (Table \ref{table_edge_measures}). Based on our analysis of edge betweenness, embeddedness and dispersion in model and real networks, there seems to be no practical advantage in using degree as node weight in the definition of the Forman curvature of an edge.

We should emphasize here that our experimental results are in concordance with a theoretical advantage of the combinatorial weights over degree, namely that, while using the second alternative, it is straightforward to produce edges having the same number of neighbours, but different Forman curvatures, thus rendering Forman curvature as a bad tool in characterization of the local topology of a network.


\subsection{Network robustness and relative importance of edge-based measures}

\begin{figure}
\includegraphics[width=.7\columnwidth]{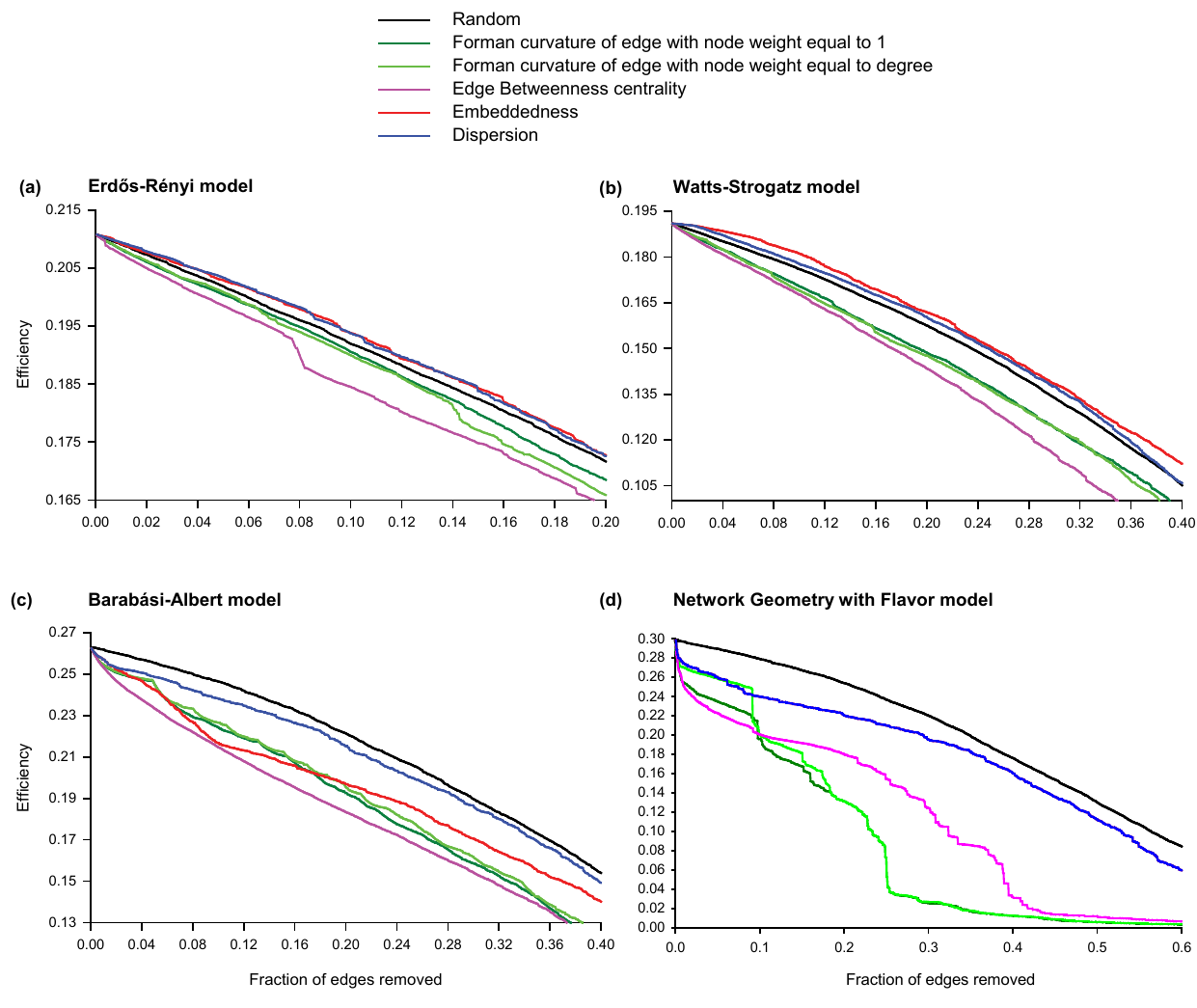}
\caption{Communication efficiency as a function of the fraction of edges removed in model networks. (a) Erd\"{o}s-R\`{e}nyi (ER) model. (b) Watts-Strogratz (WS) model. (c) Barab\`{a}si-Albert (BA) model. (d) Network geometry with flavor (NGF) model where flavor $s=1$ and dimension $d=2$. Here, the order in which the edges are removed is based on the following criteria: random order, increasing order of the Forman curvature of an edge with node weight equal to 1, increasing order of the Forman curvature of an edge with node weight equal to degree, decreasing order of edge betweenness, decreasing order of embeddedness, and decreasing order of dispersion.}
\label{rob_model_edge}
\end{figure}

\begin{figure}
\includegraphics[width=.7\columnwidth]{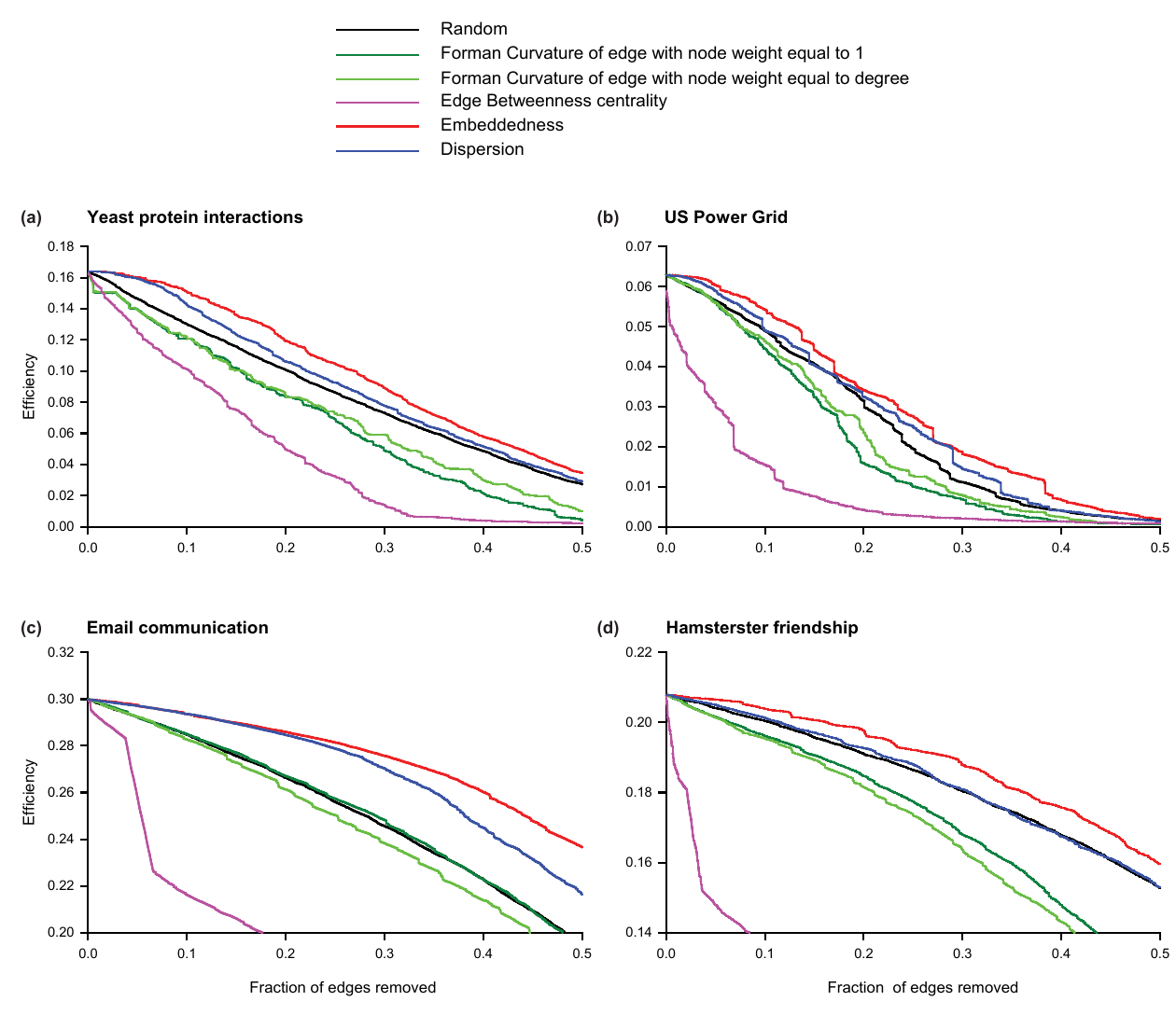}
\caption{Communication efficiency as a function of the fraction of edges removed in real networks. (a) Yeast protein interactions. (b) US Power Grid. (c) Email communication. (d) Hamsterster friendship. Here, the order in which the edges are removed is based on the same criteria as in Figure \ref{rob_model_edge}.}
\label{rob_real_edge}
\end{figure}

We next investigated the effect of removing edges based on their Forman curvature, edge betweenness, embeddedness or dispersion on the large-scale connectivity of model and real networks (Figures \ref{rob_model_edge} and \ref{rob_real_edge}). In order to quantify the large-scale connectivity of a network, we used the global network measure, communication efficiency \cite{Latora2001}, which captures a network's resilience to failure in the face of perturbations. Specifically, we have investigated the communication efficiency in model and real networks as a function of the fraction of edges removed (Figures \ref{rob_model_edge} and \ref{rob_real_edge}). In our simulations, the order of removing edges was based on the following criteria: (a) random order, (b) increasing order of the Forman curvature of an edge with node weight equal to 1, (c) increasing order of the Forman curvature of an edge with node weight equal to degree, (d) decreasing order of edge betweenness centrality, (e) decreasing order of embeddedness, and (f) decreasing order of dispersion. We find that removing edges in model and real networks based on increasing order of the Forman curvature for the two choices of node weights leads to faster disintegration compared to removing edges in random order or based on decreasing order of embeddedness or based on decreasing order of dispersion (Figures \ref{rob_model_edge} and \ref{rob_real_edge}). We find that removing edges based on decreasing order of edge betweenness centrality leads to faster disintegration of network compared to increasing order of the Forman curvature for the two choices of node weights in ER, WS and BA networks (Figure \ref{rob_model_edge}). However, in NGF networks with geometrical structure, we find that removing edges based on increasing order of the Forman curvature for the two choices of node weights leads to faster disintegration compared to decreasing order of edge betweenness centrality (Figure \ref{rob_model_edge}). Among considered real networks, we find that removing edges based on decreasing order of edge betweenness centrality leads to faster disintegration of network compared to increasing order of the Forman curvature for the two choices of node weights (Figure \ref{rob_real_edge}). Therefore, edges with highly negative Forman curvature are more important than edges with high embeddedness or high dispersion for maintaining the large-scale connectivity of networks. However, edges with highly negative Forman curvature are relatively less important than edges with high edge betweenness centrality in most networks except NGF networks with explicit geometric structure. This comes as little surprise, since betweenness centrality is a global measure which is difficult to compute for large networks, whereas Forman curvature is a local one which is much more easy to compute in large networks. Still it is interesting to find that Forman curvature, a geometrical concept, outperforms edge betweenness centrality in NGF networks with real geometric structure. Finally, between the two choices of node weights in the definition of the Forman curvature of an edge, we find that there isn't any difference between the Forman curvature with node weight equal to 1 and the Forman curvature with node weight equal to degree while determining the importance of an edge in model and real networks (Figures \ref{rob_model_edge} and \ref{rob_real_edge}).


\subsection{Distribution of normalized and unnormalized Forman curvature of nodes}

\begin{figure}
\includegraphics[width=.9\columnwidth]{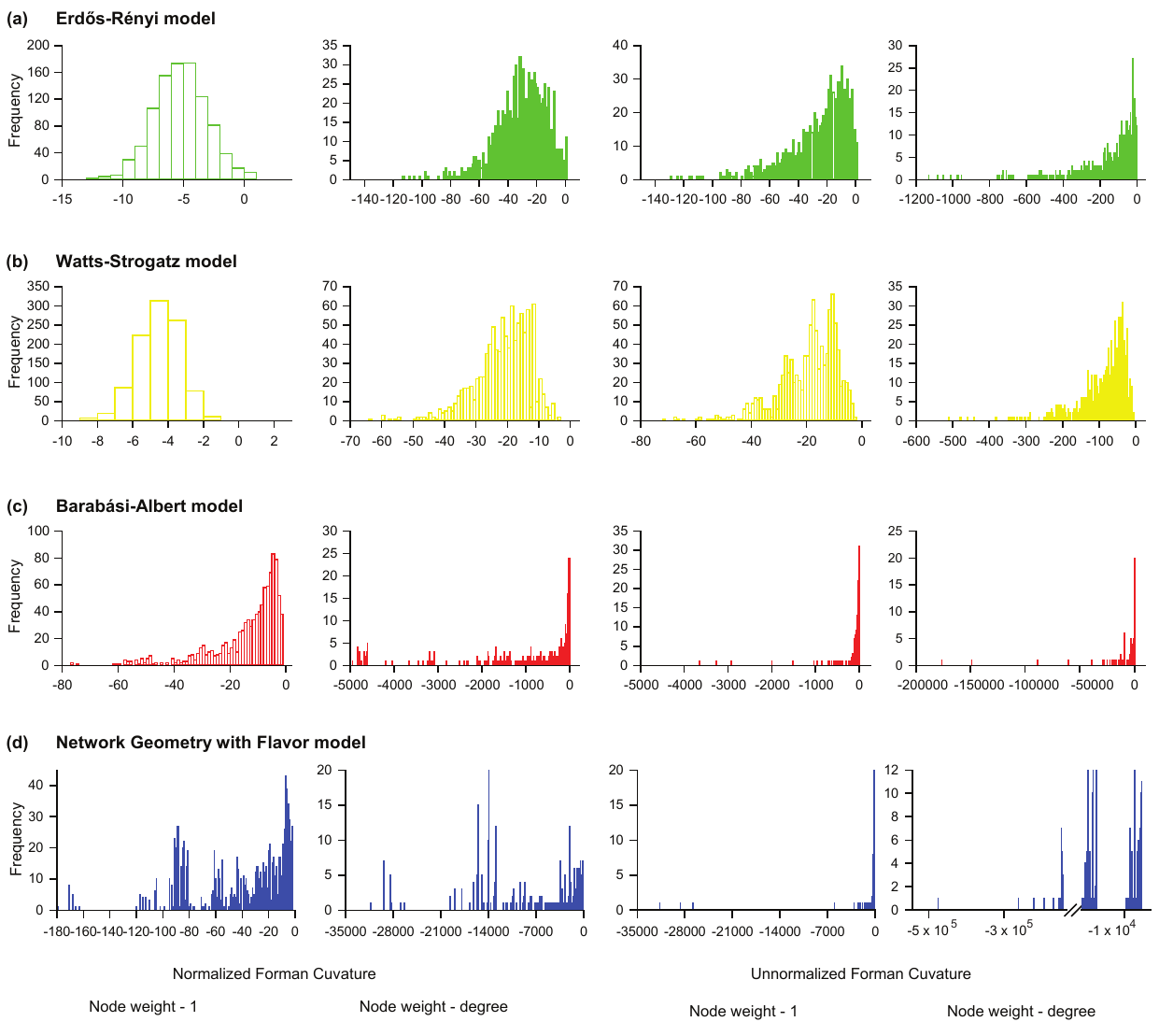}
\caption{Distribution in model networks of normalized and unnormalized Forman curvature for the two choices of node weights. (a) Erd\"{o}s-R\`{e}nyi (ER) model. (b) Watts-Strogratz (WS) model. (c) Barab\`{a}si-Albert (BA) model. (d) Network geometry with flavor (NGF) model where flavor $s=1$ and dimension $d=2$.}
\label{dist_model_node}
\end{figure}

\begin{figure}
\includegraphics[width=.9\columnwidth]{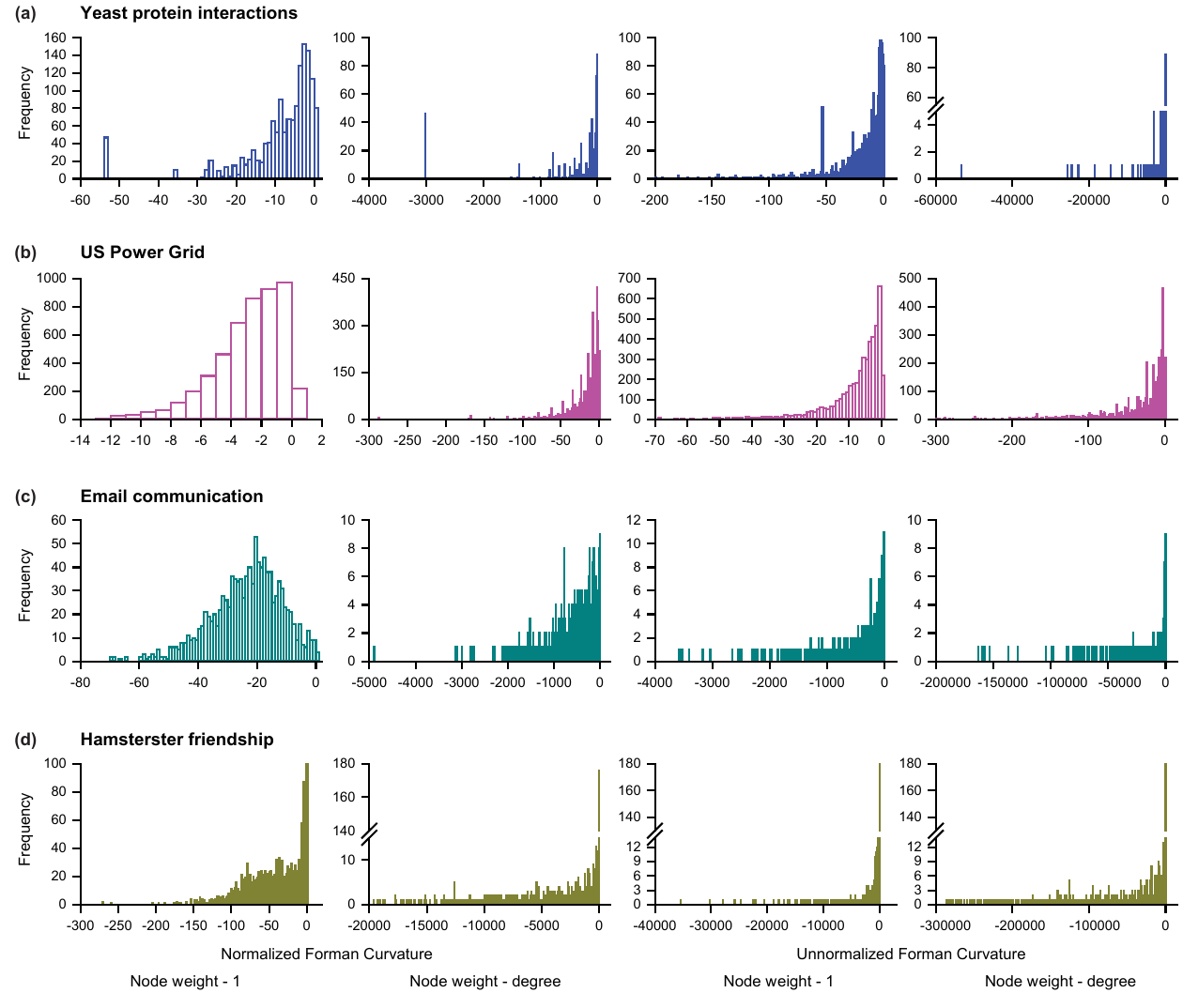}
\caption{Distribution in real networks of normalized and unnormalized Forman curvature for the two choices of node weights. (a) Yeast protein interactions. (b) US Power Grid. (c) Email communication. (d) Hamsterster friendship.}
\label{dist_real_node}
\end{figure}

We next investigated the distribution of normalized and unnormalized Forman curvature of nodes in model and real networks (Figures \ref{dist_model_node} and \ref{dist_real_node}). Obviously, the magnitude of unnormalized node curvature is much higher than normalized node curvature, and the magnitude of curvature with node weight equal to degree is higher than curvature with node weight equal to 1 (Figures \ref{dist_model_node} and \ref{dist_real_node}). Among model networks, at a qualitative level, the nature of distribution of both normalized and unnormalized Forman curvature of nodes is same in ER, WS and BA networks (Figure \ref{dist_model_node}). Specifically, we find that the distribution in narrow for both normalized and unnormalized Forman curvature of nodes in random ER networks and small-world WS networks, while the distribution is broad for both normalized and unnormalized Forman curvature of nodes in scale-free BA networks (Figure \ref{dist_model_node}). In NGF network with $s=1$ and $d=2$, the distribution is broad for both normalized and unnormalized Forman curvature of nodes though there are some clear differences in the shape of each distribution (Figure \ref{dist_model_node}). These and our previous results suggest that both normalized \cite{Sreejith2016} and unnormalized definitions of Forman curvature of nodes can be used to distinguish between different types of model networks. Among real networks, at a qualitative level, the nature of distribution of normalized and unnormalized Forman curvature of nodes can be different for a given real network (Figure \ref{dist_real_node}). For example, the nature of distribution of normalized Forman curvature with node weight equal to 1 in the Email communication network is different from normalized Forman curvature with node weight equal to degree or unnormalized Forman curvature for the two choices of node weights (Figure \ref{dist_real_node}(c)). However, similar to scale-free BA networks, the distribution of both normalized and unnormalized Forman curvature of nodes is broad in considered real networks (Figure \ref{dist_real_node}).


\subsection{Association of node-based measures with normalized and unnormalized Forman curvature of nodes}

\begin{figure}
\includegraphics[width=.9\columnwidth]{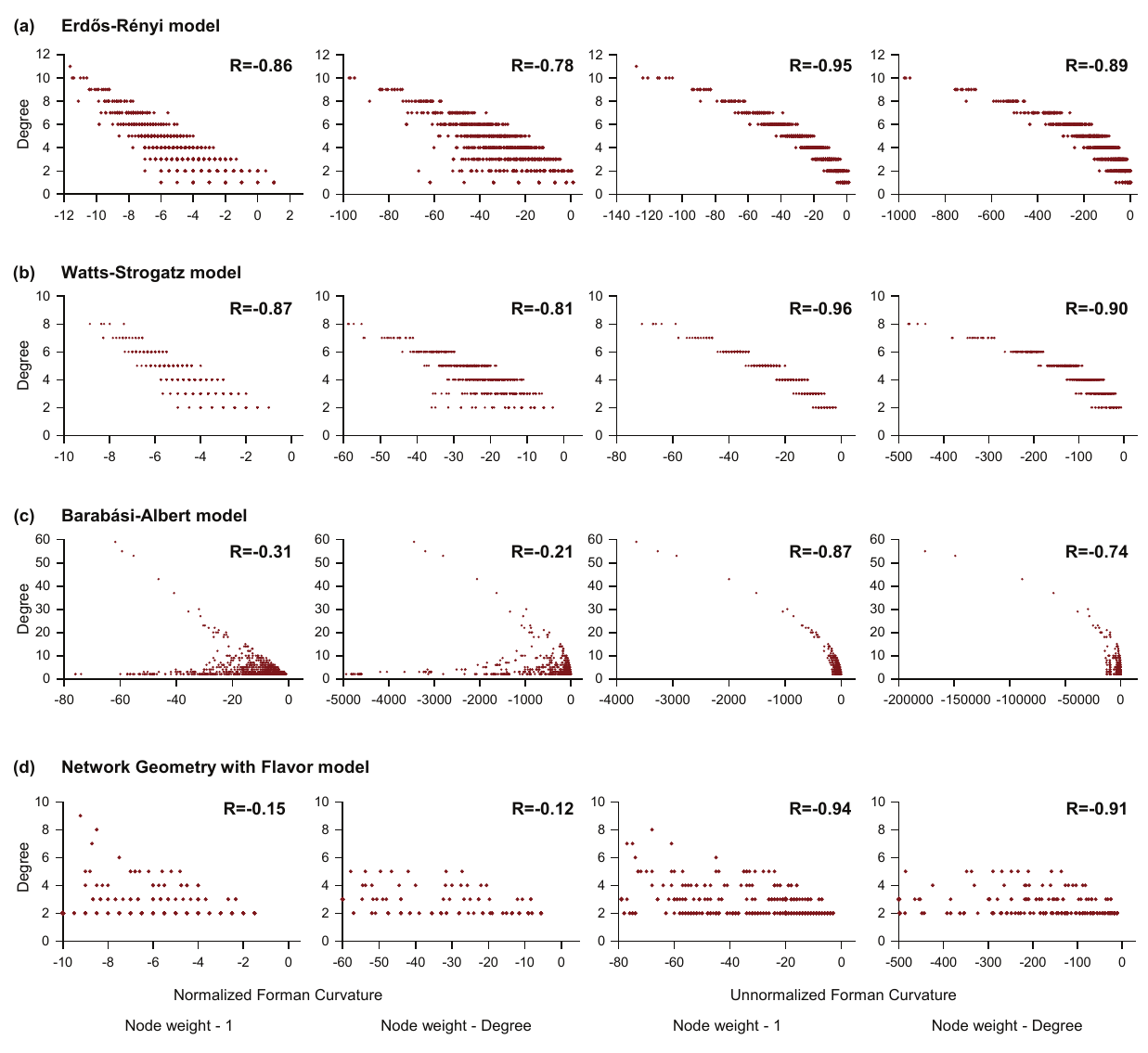}
\caption{Correlation between node degree and normalized or unnormalized Forman curvature for the two choices of node weights in model networks. (a) Erd\"{o}s-R\`{e}nyi (ER) model. (b) Watts-Strogratz (WS) model. (c) Barab\`{a}si-Albert (BA) model. (d) Network geometry with flavor (NGF) model where flavor $s=1$ and dimension $d=2$. We indicate the Pearson correlation coefficient $R$ in each plot.}
\label{corr_model_degree_node}
\end{figure}

\begin{figure}
\includegraphics[width=.9\columnwidth]{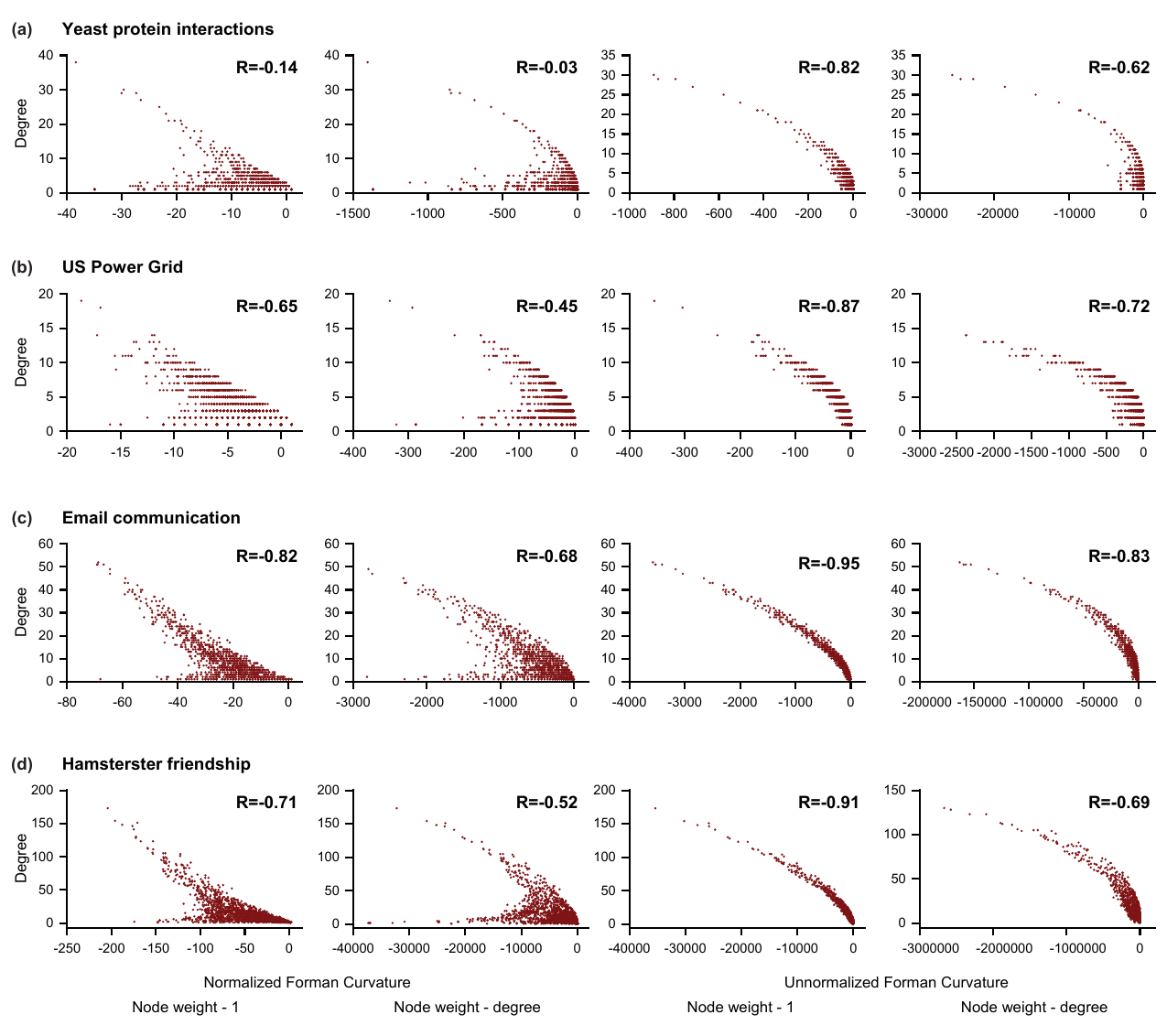}
\caption{Correlation between node degree and normalized or unnormalized Forman curvature for the two choices of node weights in real networks. (a) Yeast protein interactions. (b) US Power Grid. (c) Email communication. (d) Hamsterster friendship. We indicate the Pearson correlation coefficient $R$ in each plot.}
\label{corr_real_degree_node}
\end{figure}

Node degree gives the number of edges incident to that node in a network. We have investigated the correlation between degree and normalized or unnormalized Forman curvature of nodes in model and real networks (Figures \ref{corr_model_degree_node} and \ref{corr_real_degree_node}; Supplementary Table S1). Among the model networks, we find that both normalized and unnormalized Forman curvature (for the two choices of node weights) have high negative correlation with node degree in random ER networks and small-world WS networks (Figure \ref{corr_model_degree_node}; Supplementary Table S1). However, in scale-free BA networks and NGF networks with geometric structure, the unnormalized Forman curvature has much higher negative correlation with node degree compared to the normalized Forman curvature (Figure \ref{corr_model_degree_node}; Supplementary Table S1). We highlight that the unnormalized Forman curvature with node weight equal to 1 has the highest negative correlation with node degree in considered model networks (Figure \ref{corr_model_degree_node}; Supplementary Table S1). Among the real networks, we also find that the unnormalized Forman curvature with node weight equal to 1 has the highest negative correlation with node degree (Figure \ref{corr_real_degree_node}; Supplementary Table S1). Surprisingly, we find no correlation between normalized Forman curvature and node degree in the Facebook network while a high negative correlation is obtained between unnormalized Forman curvature and node degree in the same network (Supplementary Table S1). Note that between the two possible choices of node weights in the definition of the Forman curvature of an edge, we find that the choice of node weight equal to 1 leads to highest negative correlation between unnormalized Forman curvature and node degree in model and real networks (Supplementary Table S1).

\begin{figure}
\includegraphics[width=.9\columnwidth]{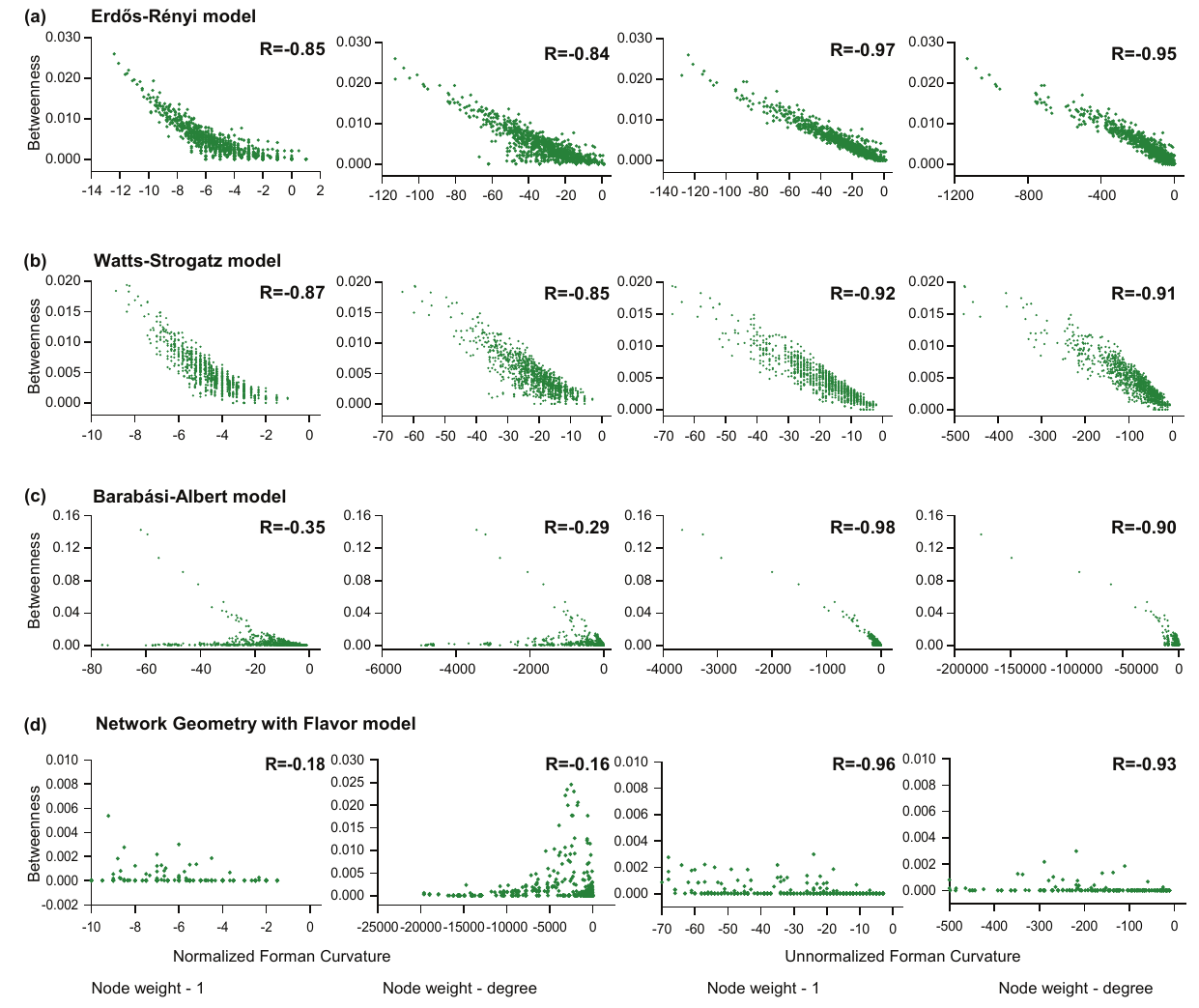}
\caption{Correlation between node betweenness centrality and normalized or unnormalized Forman curvature for the two choices of node weights in model networks. (a) Erd\"{o}s-R\`{e}nyi (ER) model. (b) Watts-Strogratz (WS) model. (c) Barab\`{a}si-Albert (BA) model. (d) Network geometry with flavor (NGF) model where flavor $s=1$ and dimension $d=2$. We indicate the Pearson correlation coefficient $R$ in each plot.}
\label{corr_model_bet_node}
\end{figure}

\begin{figure}
\includegraphics[width=.9\columnwidth]{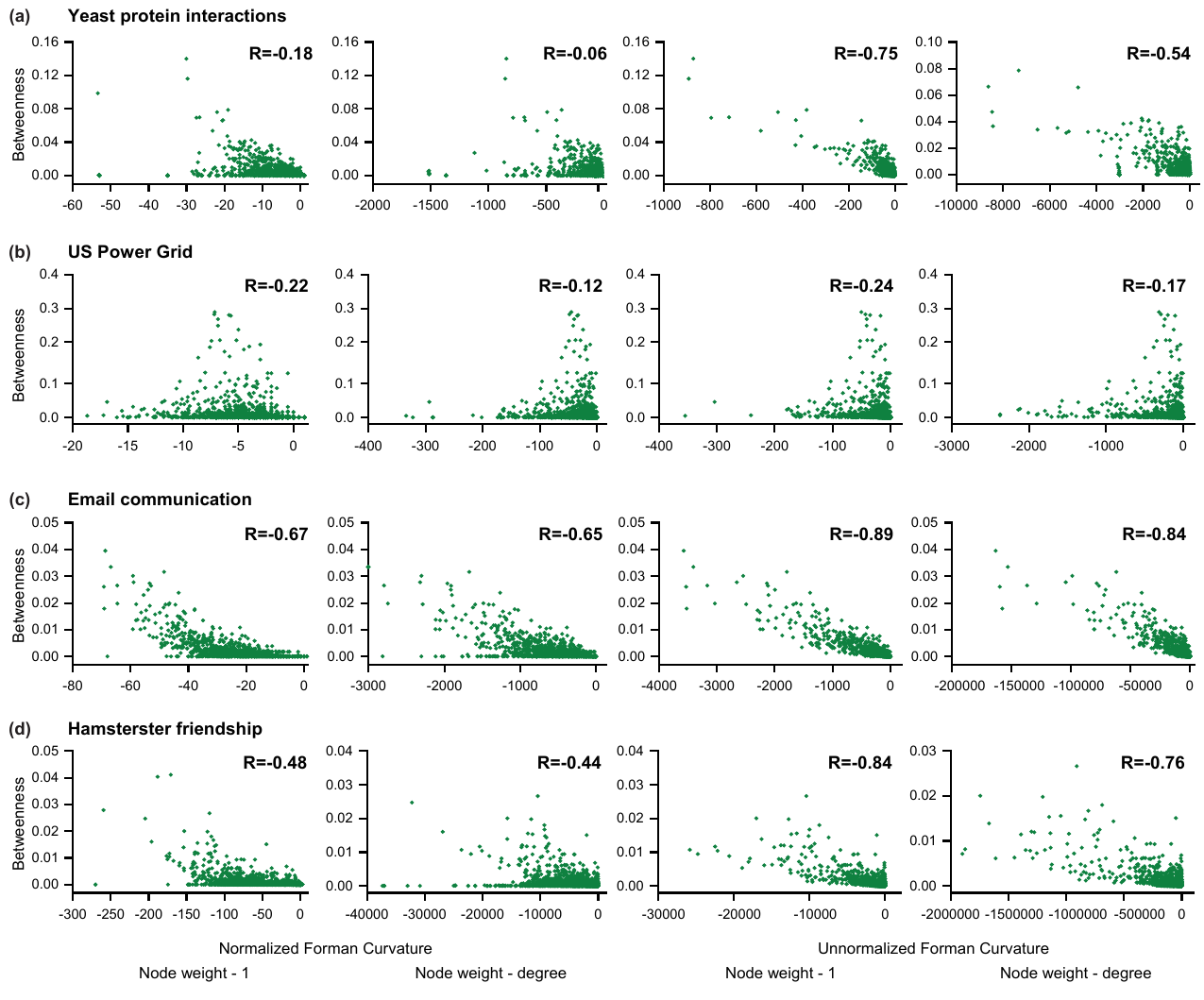}
\caption{Correlation between node betweenness centrality and normalized or unnormalized Forman curvature for the two choices of node weights in real networks. (a) Yeast protein interactions. (b) US Power Grid. (c) Email communication. (d) Hamsterster friendship. We indicate the Pearson correlation coefficient $R$ in each plot.}
\label{corr_real_bet_node}
\end{figure}

Node betweennness centrality \cite{Freeman1977,Newman2010} quantifies the fraction of shortest paths between all pairs of nodes in the network that pass through that node. We have also investigated the correlation between betweenness centrality and normalized or unnormalized Forman curvature of nodes in model and real networks (Figures \ref{corr_model_bet_node} and \ref{corr_real_bet_node}; Supplementary Table S1). We find that the unnormalized Forman curvature with node weight equal to 1 has the highest negative correlation with betweenness centrality in considered model networks (Figure \ref{corr_model_bet_node}; Supplementary Table S1). Among the real networks, we also find that the unnormalized Forman curvature with node weight equal to 1 has the highest negative correlation with betweenness centrality in the majority of real networks (Figure \ref{corr_real_bet_node}; Supplementary Table S1). Jazz musicians network is the only exception where the unnormalized Forman curvature with node weight equal to degree has slightly higher negative correlation with betweenness centrality compared to the unnormalized Forman curvature with node weight equal to 1 (Supplementary Table S1). These results underscore that the unnormalized Forman curvature with node weight equal to 1 is in practice better for the analysis of model and real networks.

The clustering coefficient \cite{Holland1971,Watts1998} of a node quantifies the number of edges that are realized between the neighbours of the node divided by the number of edges that could possibly exist between the neighbours of the node in the network. The clustering coefficient has been proposed as a measure to quantify the curvature of networks \cite{Eckmann2002}. We analyzed the correlation between clustering coefficient and normalized or unnormalized Forman curvature of nodes in model and real networks (Supplementary Table S1). We find that both normalized and unnormalized Forman curvature have no or weak correlation with clustering coefficient in model and real networks (Supplementary Table S1). Thus, Forman curvature of nodes has better association with degree and centrality measures than clustering coefficient in complex networks. These findings extend our previous work \cite{Sreejith2016} where no or weak correlation was found between the normalized Forman curvature with node weight equal to 1 and clustering coefficient of nodes in model and real networks.


\subsection{Network robustness and relative importance of normalized and unnormalized Forman curvature of nodes}

\begin{figure}
\includegraphics[width=.7\columnwidth]{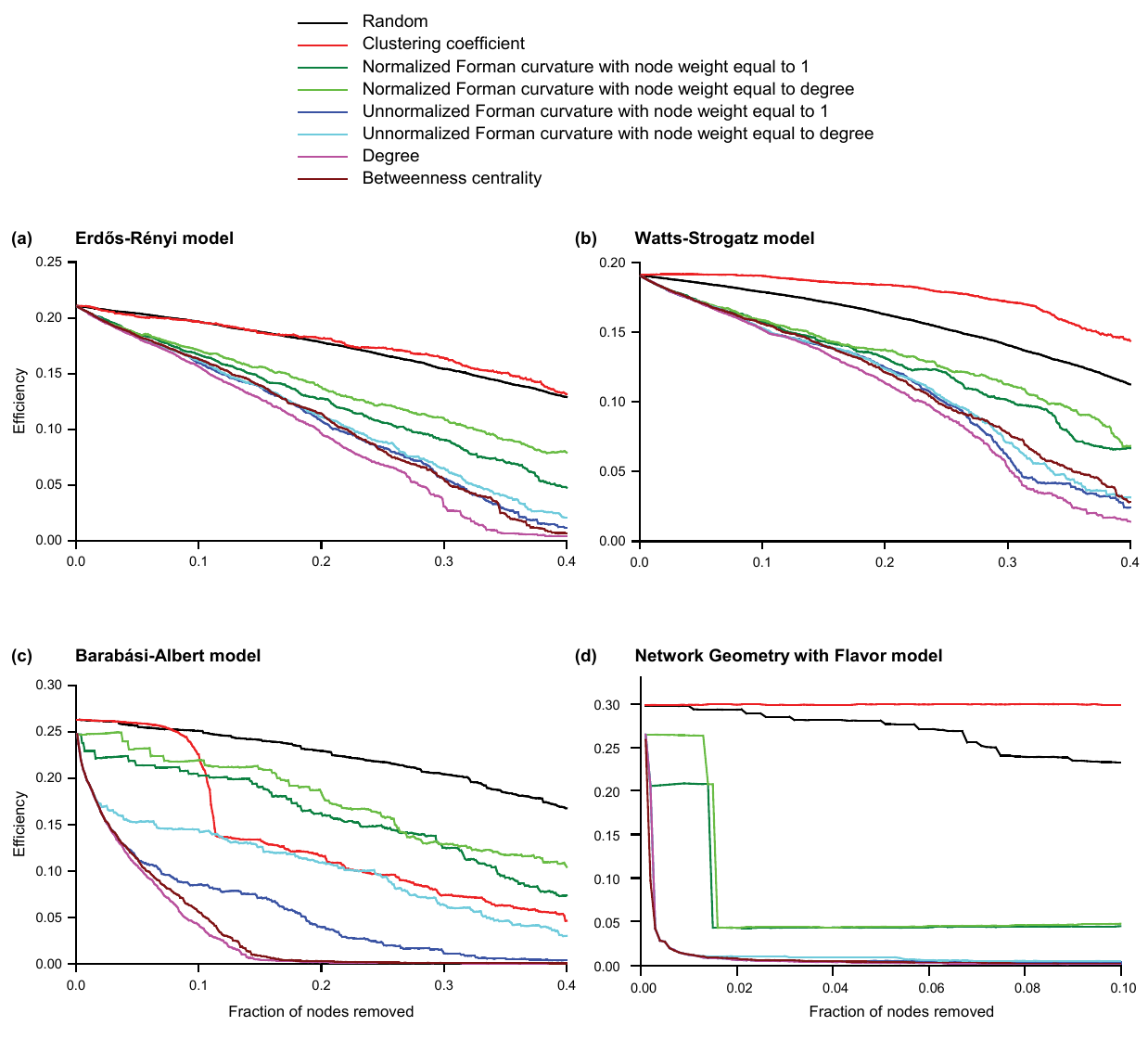}
\caption{Communication efficiency as a function of the fraction of nodes removed in model networks. (a) Erd\"{o}s-R\`{e}nyi (ER) model. (b) Watts-Strogratz (WS) model. (c) Barab\`{a}si-Albert (BA) model. (d) Network geometry with flavor (NGF) model where flavor $s=1$ and dimension $d=2$. Here, the order in which the nodes are removed is based on the following criteria: random order, decreasing order of clustering coefficient, increasing order of the normalized Forman curvature with node weight equal to 1, increasing order of the normalized Forman curvature with node weight equal to degree, increasing order of the unnormalized Forman curvature with node weight equal to 1, increasing order of the unnormalized Forman curvature with node weight equal to degree, decreasing order of degree, and decreasing order of betweenness centrality.}
\label{rob_model_node}
\end{figure}

\begin{figure}
\includegraphics[width=.7\columnwidth]{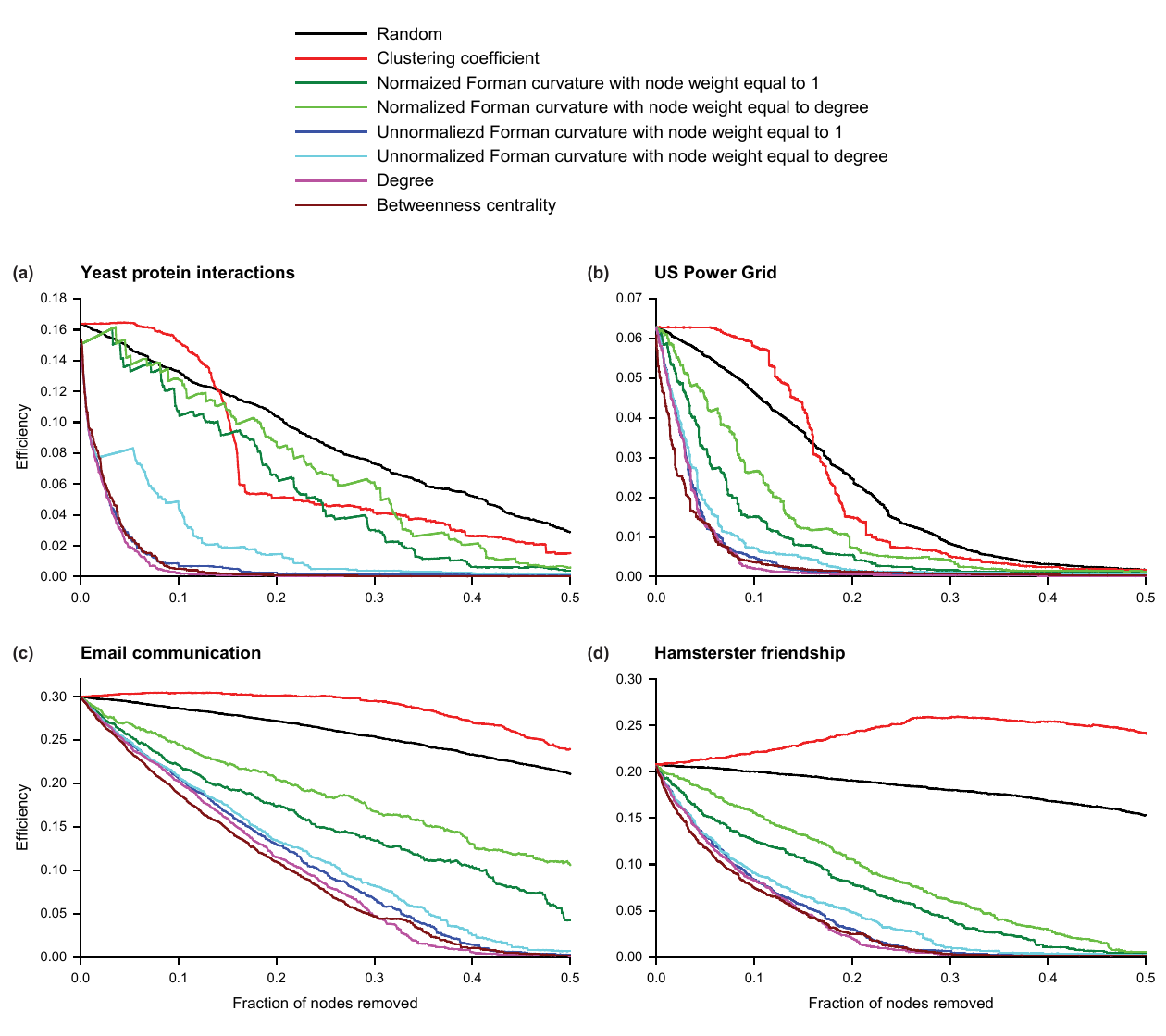}
\caption{Communication efficiency as a function of the fraction of nodes removed in real networks. (a) Yeast protein interactions. (b) US Power Grid. (c) Email communication. (d) Hamsterster friendship. Here, the order in which the nodes are removed is based on the same criteria as in Figure \ref{rob_model_node}.}
\label{rob_real_node}
\end{figure}

We next investigated the effect of removing nodes based on normalized or unnormalized Forman curvature on the communication efficiency of model and real networks (Figures \ref{rob_model_node} and \ref{rob_real_node}). In our simulations, the order of removing nodes was based on the following criteria: (a) random order, (b) decreasing order of clustering coefficient, (c) increasing order of normalized Forman curvature with node weight equal to 1, (d) increasing order of normalized Forman curvature with node weight equal to degree, (e) increasing order of unnormalized Forman curvature with node weight equal to 1, (f) increasing order of unnormalized Forman curvature with node weight equal to degree, (g) decreasing order of degree, and (f) decreasing order of betwenness centrality. Previously we \cite{Sreejith2016} have shown that removing nodes in model and real networks based on increasing order of normalized Forman curvature with node weight equal to 1 leads to faster disintegration compared to removing nodes in random order or based on decreasing order of clustering coefficient. However, we had also found that removing nodes in model and real networks based on decreasing order of degree or decreasing order of betweenness centrality typically leads to faster disintegration compared to increasing order of normalized Forman curvature with node weight equal to 1. Thus, nodes with highly negative normalized Forman curvature with node weight equal to 1 were found to be more important than nodes with high clustering coefficient but relatively less important than nodes with high degree or high betweenness centrality for the large-scale connectivity of networks. Here, we extended our previous work \cite{Sreejith2016} to show that removing nodes in model and real networks based on increasing order of normalized or unnormalized Forman curvature leads to faster disintegration compared to removing nodes in random order or based on decreasing order of clustering coefficient (Figures \ref{rob_model_node} and \ref{rob_real_node}). Furthermore, we show that removing nodes in model and real networks based on increasing order of unnormalized Forman curvature leads to even faster disintegration of network compared to increasing order of normalized Forman curvature (Figures \ref{rob_model_node} and \ref{rob_real_node}). Lastly, we find that removing nodes in model and real networks based on increasing order of unnormalized Forman curvature with node weight equal to 1 has close to similar effect of removing nodes based on decreasing order of degree or decreasing order of betweenness centrality in most networks (Figures \ref{rob_model_node} and \ref{rob_real_node}). These results underscore that unnormalized Forman curvature (especially, the one with node weight equal to 1) is a better indicator of the importance of a node compared to normalized Forman curvature in model and real networks.


\section{Conclusions}
\label{conclusion}

In summary, we have undertaken a systematic evaluation of a new combinatorial curvature -- Forman curvature -- in model and real networks. Forman curvature is an edge-based measure which is simple to compute in large networks. In this contribution, we have compared Forman curvature with other edge-based measures, edge betweenness, embeddedness and dispersion, in model and real networks. In model networks, Forman curvature of edges has significant negative correlation with edge betweenness, no correlation or moderate negative correlation with embeddedness and dispersion. In real networks, Forman curvature of edges has no correlation or weak correlation with edge betweenness, no correlation to significant negative correlation with embeddedness, and mostly no correlation with dispersion. Moreover, Forman curvature of an edge in model and real networks is a superior indicator of the importance of edges in comparison to embeddedness or dispersion
but is an inferior indicator in comparison to edge betweennes. An exception is NGF networks with explicit geometric structure where Forman curvature of an edge is a superior indicator of the importance of edges in comparison to edge betweennes. This suggests that Forman curvature, a geometric notion, may be best suited to analyze networks with real geometrical structure.

While computing the Forman curvature of an edge in networks without prescribed weights, there are at least two possible choices for node weights, combinatorial weight equal to 1 or weight equal to degree. While defining the Forman curvature of nodes in networks, there are two possible choices, normalized Forman curvature or unnormalized Forman curvature. We have investigated the relative advantages of these fundamental choices in the implementation of Forman curvature in model and real networks. We find that the unnormalized Forman curvature of nodes has much higher negative correlation with degree and centrality measures compared to normalized Forman curvature in model and real networks. Moreover, the unnormalized Forman curvature is a better indicator of the importance of nodes in model and real networks. Moving on to the two natural choices for node weights while computing the Forman curvature, we remark that node degree is intrinsic in the definition of the Forman curvature of an edge as it appears implicitly in the sum over adjacent edges in the defining formula (Eq. \ref{FormanRicciEdge}). Thus, the inclusion of node degree as node weight in the definition of the Forman curvature could be redundant. Still, the second choice of node degree as node weight seems definitely a legitimate option, and a comparative analysis of both choices in model and real networks was performed here to determine the better alternative. Our empirical analysis of Forman curvature finds that there is no practical advantage in using degree as node weight. Thus, even on the practical level, the fundamental theoretical choice of combinatorial weights is the preferable option. Thus, we find that unnormalized Forman curvature with combinatorial node weight equal to 1 is the preferred option in practice for the analysis of complex networks. As the unnormalized Forman curvature of a node is analogous to scalar curvature in Riemannian geometry, one might also conclude from these empirical observations that the geometry inspired definition of node curvature is more powerful when dealing with the Forman's discretization of the classical Ricci curvature, a geometric notion.


\section*{Acknowledgments}
\noindent We would like to thank the anonymous reviewer for insightful suggestions which have helped improve the manuscript. We would also like to thank M. Karthikeyan and R.P. Vivek-Ananth for help with figures. AS acknowledges support from Max Planck Society, Ramanujan fellowship (SB/S2/RJN-006/2014) and Department of Science and Technology (DST) Start-up project (YSS/2015/000060).

\section*{Supplementary Information}
\noindent \textbf{Supplementary Table S1:} Association of node-based measures with normalized and unnormalized Forman curvature of nodes in model and real networks.												

\section*{References}


\pagestyle{empty}
\begin{landscape}
\begin{table}
\caption{Association of edge-based measures with the Forman curvature of edges in model and real networks.}
\label{table_edge_measures}
\begin{tabular}{|l|c|c|c|c|c|c|}
\hline
\multirow{3}{*}{\textbf{\small Network}} & \multicolumn{3}{c|}{\textbf{\small Forman curvature with}} & \multicolumn{3}{c|}{\textbf{\small Forman curvature with}} \\
& \multicolumn{3}{c|}{\textbf{\small node weight equal to 1}} & \multicolumn{3}{c|}{\textbf{\small node weight equal to degree}} \\
\cline{2-7}
& \textbf{\small Edge betweenness} & \textbf{\small Embeddedness} & \textbf{\small Dispersion}  & \textbf{\small Edge betweenness} & \textbf{\small Embeddedness} & \textbf{\small Dispersion} \\
\hline
\textbf{Model networks} & & & & & &              \\
\small{Erd\"{o}s-R\`{e}nyi (ER) model with $p=0.004$}  & -0.78  & -0.08 & 0.0   & -0.77 & -0.08 & 0.0   \\
\small{Watts-Strogratz (WS) model with $k=5$ and $p=0.5$}      & -0.71  & -0.10 & -0.08 & -0.71 & -0.10 & -0.08 \\
\small{Barab\`{a}si-Albert (BA) model with $m=2$}  & -0.68  & -0.47 & -0.28 & -0.68 & -0.47 & -0.28 \\
\small{ Network geometry with dimension $d=2$ and flavor $s=-1$} & -0.50 & -0.59 & -0.59 & -0.56 & -0.44 & -0.44 \\
\small{ Network geometry with dimension $d=2$ and flavor $s=0$}  & -0.30 & -0.30 & -0.31 & -0.24 & -0.20 & -0.22 \\
\small{ Network geometry with dimension $d=2$ and flavor $s=1$}  & -0.35 & -0.26 & -0.20 & -0.31 & -0.21 & -0.16 \\
\small{ Network geometry with dimension $d=2$ and flavor $s=-1$} & -0.36 & -0.43 & -0.40 & -0.27 & -0.26 & -0.28 \\
\small{ Network geometry with dimension $d=2$ and flavor $s=0$}  & -0.46 & -0.29 & -0.20 & -0.40 & -0.21 & -0.16 \\
\small{ Network geometry with dimension $d=2$ and flavor $s=1$}  & -0.36 & -0.23 & -0.16 & -0.30 & -0.17 & -0.14 \\
\hline
\textbf{Real networks} & & & & & &               \\
\small{Adjective-Noun adjacency}   & -0.38  & -0.76  & -0.50  & -0.46 & -0.69 & -0.51  \\
\small{Email communication}        & -0.29  & -0.48  & -0.38  & -0.35 & -0.40 & -0.37  \\
\small{Euro Road}                  & -0.22  & -0.32  & -0.10  & -0.20 & -0.25 & -0.09  \\
\small{Facebook}                   & -0.03  & 0.07   & 0.01   & -0.02 & 0.08 & 0.02    \\
\small{Gnutella}                   & 0.04   & -0.58  & -0.30  & 0.09 & -0.60 & -0.32   \\
\small{Hamsterster friendship}     & -0.24  & -0.44  & -0.10  & -0.25 & -0.31 & -0.11  \\
\small{Human protein interactions} & -0.42  & -0.17  & -0.15  & -0.31 & -0.08 & -0.12  \\
\small{Jazz musicians}             & -0.10  & -0.67  & -0.19  & -0.23 & -0.54 & -0.23  \\
\small{PDZ domain interactions}    & -0.34  & 0.02   & 0.0    & -0.23 & 0.05 & 0.0     \\
\small{US Power Grid}              & -0.12  & -0.49  & -0.24  & -0.06 & -0.40 & -0.23  \\
\small{Yeast protein interactions} & -0.34  & -0.08  & -0.08  & -0.19 & 0.04 & -0.02   \\
\hline
\end{tabular}
\end{table}
\end{landscape}

\end{document}